\input harvmac
\overfullrule=0pt
\def\Title#1#2{\rightline{#1}\ifx\answ\bigans\nopagenumbers\pageno0\vskip1in
\else\pageno1\vskip.8in\fi \centerline{\titlefont #2}\vskip .5in}

\lref\romans{L. Romans, Nucl. Phys. {\bf B276} (1986) 71.} 
\lref\kimn{H. Kim, L. Romans and P. Nieuwenhuizen, Phys. Rev. {\bf D32} (1985)
389}
\lref\vw{C. Vafa and E. Witten,
 Nucl. Phys. {\bf B431} (1994) 3,
 hep-th/9408074.}
\lref\gbody{everybody}
\lref\clp{S.\ Carlip, Phys. Rev. {\bf D51} (1995) 632.}
\lref\scct{S. Carlip and C. Teitelboim,}
\lref\ascv{A.\ Strominger and C.\ Vafa,
 { Phys. Lett.} {\bf B379}
(1996) 99, hep-th/9601029.}
\lref\nhbh{A. Strominger, hep-th/9712251.}
\lref\ho{G. Horowitz and H.Ooguri, hep-th/9802116.}
\lref\dps{M.\ Douglas, J.\
Polchinski and A.\ Strominger, hep-th/9703031.}
\lref\jm{J.\ Maldacena,
hep-th/9711200.}
\lref\sh{S.\ Hyun, hep-th/9704005.}
\lref\bph{H.\ Boonstra, B.\ Peeters, and K.\ Skenderis, hep-th/9706192.}
\lref\ss{K.\ Sfetsos and K.\ Skenderis, hep-th/9711138.}
\lref\btz{M.\ Banados, C.\ Teitelboim and J.\ Zanelli, {\sl
Phys. Rev. Lett.} {\bf 69} (1992) 1849.}
\lref\bhtz{M.\ Banados, M.\ Henneaux, C.\ Teitelboim, and J.\ Zanelli, 
{Phys.Rev.} {\bf D48} (1993) 1506, gr-qc/9302012.}
\lref\ls{D. A . Lowe and A. Strominger, Phys. Rev. Lett {\bf 73} (1994) 1468.}
\lref\jc{J. A. Cardy, Nucl. Phys. {\bf B270} (1986) 186.}
\lref\kal{P. Claus, R. Kallosh, J. Kumar, P. Townsend and A. van Proeyen,
hep-th/9801206.}
\lref\scct{S. Carlip and C. Teitelboim, Phys. Rev. {\bf D51} (1995) 622. }
\lref\nhg{M. Cvetic and A. Tseytlin, Phys. Rev. {\bf D53} (1996) 5619.}
\lref\hw{G. T. Horowitz and D. Welch, Phys. Rev. Lett. {\bf 71} (1993) 328.}
\lref\chv{O.\ Coussaert, M.\ Henneaux and P.\ van Driel, Class. 
Quant. Grav {\bf 12} (1995) 2961.}
\lref\ch{O.\ Coussaert and M.\ Henneaux,
 Phys. Rev. Lett {\bf 72} (1994) 183, hep-th/9310194.}
\lref\nhg{M. Cvetic and A. Tseytlin, Phys. Rev. {\bf D53} (1996) 5619.}
\lref\chtwo{O.\ Coussaert and M.\ Henneaux, hep-th/9407181.}
\lref\rez{B. Reznik, Phys. Rev. {\bf D51} (1995) 1728.}
\lref\grta{D. J. Gross, Nucl. Phys. {\bf B400} (1993) 161}
\lref\jbmh{J.\ D.\ Brown and M.\
Henneaux, { Comm. Math. Phys.} {\bf 104} (1986) 207.}
\lref\hms{G. Horowitz, J. Maldacena and A. Strominger,
Phys. Lett. {\bf B383} (1996) 151, hep-th/9603109.  }
\lref\ofer{ O. Aharony, M. Berkooz and N. Seiberg,
 hep-th/9712117.}
\lref\withol{E. Witten,
 hep-th/9802150.}
\lref\stabil{P. Breitenlohner and D. Freedman,
Phys. Lett. {\bf 115 B} (1982) 197; 
Ann. Phys. {\bf 114} (1982) 197.}
\lref\steif{A. Steif, Phys.Rev. {\bf D53} (1996) 5521.}
\lref\cang{D. Cangemi, M. LeBlanc and R. Mann, Phys. Rev.  {\bf D48} 
(1993) 3606.}
\lref\mtrx{J. Polchinski, Nucl.Phys. {\bf B362} (1991) 125. }
\lref\msu{
J. Maldacena and L. Susskind, 
Nucl.Phys. {\bf B475} (1996) 679, hep-th/9604042.}

\lref\vafaud{C. Vafa,
Nucl. Phys. {\bf B463 } (1996) 435,
hep-th/9512078.}
\lref\gkp{S.S. Gubser, I.R. Klebanov, A.M. Polyakov,
hep-th/9802109.}
\lref\jmas{J. Maldacena and A. Strominger, Phys.Rev. {\bf D55} (1997) 861.} 
\lref\skend{ H. J. Boonstra, B. Peeters and  K. Skenderis,
hep-th/9803231. }
\lref\adf{L. Andrianopoli, R. D'Auria and S. Ferrara, Int. J. Mod. Phys.
{\bf A12} (1997) 3759, hep-th/9612105.}
\lref\kalop{N. Kaloper, Phys. Rev. {\bf D48} (1993) 2598.}
\lref\gps{S. Gubser, hep-th/9704195}
\lref\kum{A. Ali and A. Kumar, Mod. Phys. Lett. {\bf A8} (1993) 2045.}

\def\ads {$AdS_3$}
\Title{\vbox{\baselineskip12pt
\hbox{hep-th/9804085}\hbox{HUTP-98/A016}}}
{\vbox{\centerline {AdS$_3$ Black Holes and a }
\centerline {Stringy Exclusion Principle} }}
\centerline{Juan Maldacena and Andrew Strominger}

\bigskip\centerline{Department of Physics}
\centerline{Harvard University}\centerline{Cambridge, MA 02138}
%
%
\def\co{{k }}
\def\mf{$AdS_3\times S^3\times M^4$}
\def\qof{{k }}

\def\s{\sigma}
\def\e{\epsilon}

\def\[{\left [}
\def\]{\right ]}
\def\({\left (}
\def\){\right )}
\def\ads {$AdS_3$}
\def\sll{$SL(2,R)_L$}
\def\slr{$SL(2,R)_R$}
\def\slt{$SL(2,R)_L\otimes SL(2,R)_R$}
\def\apm{{\alpha^\prime}}
\def\slc{$SL(2,C)$}
\def\p{\partial}
\vskip .3in

\centerline{\bf Abstract}
The duality relating near-horizon microstates of  
black holes obtained as orbifolds of a subset of \ads\
to the states of a conformal field theory is 
analyzed in detail. The \slt\ invariant vacuum on \ads\ corresponds  
to the NS-NS vacuum of the conformal field theory. The 
effect of the orbifolding is to produce a density matrix, the 
temperature and entropy of which coincide with the black hole. 
For string theory examples the spectrum of chiral 
primaries agrees with the 
spectrum of multi-particle BPS states for particle numbers less than  
of order the central charge. An upper bound 
on the BPS particle number follows from  the upper bound on the $U(1)$
charge of chiral primaries. This is a stringy exclusion principle 
which cannot be seen in perturbation theory 
about \ads .

\smallskip
\noindent
\Date{}
\newsec{Introduction}
 
String theory leads to a surprising and computationally powerful 
synthesis between gravity in (D+1)-dimensional 
anti-deSitter space and conformal field theory in D dimensions \jm. 
The exact duality  discovered in 
\jm\ relating 
string theory \mf , where $M^4$ is $K3$ or $T^4$,  and  
two-dimensional conformal field theory on a symmetric product of 
$M^4$, provides a rich but 
highly tractable example (investigated earlier in 
\refs{\dps \jmas \sh \ss -\kal } ) which shall be the focus of this paper. 
The richness follows in part from the fact that black holes can 
be obtained as orbifolds of a subset of \ads\ \bhtz. Both sides of 
the equation are also relatively tractable theories: the  
conformal field
theory in question is hyperkahler, while on the string theory 
side one encounters $SL(2,R)$ and $SU(2)$ WZW models. This 
contrasts with the higher dimensional cases, which require 
knowledge of higher dimensional conformal field theories and 
strings propagating in  RR backgrounds.   

In the three-dimensional case the duality between conformal field theory
and quantum gravity on \ads , as well as the central 
charge of the conformal field theory, was derived some time 
ago in a more general context 
from the observation that the asymptotic symmetry 
group of \ads\ is generated by left and right Virasoro algebras \jbmh. 
The Hilbert space must provide a representation of the algebra and
so is that of a conformal field theory. This turns out to be enough 
information to show that the number of states grows at large energy 
exactly as expected from the Bekenstein-Hawking entropy formula \nhbh.
One would like to go beyond this and explicitly identify the 
near-horizon microstates responsible for the entropy. 

In this paper 
we begin the process of  identifying 
these microstates in the case of type II string theory on
orbifolds of 
\mf. 
Since \ads\ is $SL(2,R)$ and $S^3$ is $SU(2)$, the six dimensional 
part is string theory on a group manifold \refs{\hw \kalop \kum -\nhg}. 
This is the near-horizon 
geometry of the black hole studied in \ascv\ wherein
the black hole was shown to be described by
conformal field theory whose target space is a deformation of a
 symmetric product of 
copies of $M^4$. We find that the chiral primaries in this conformal 
field theory correspond to multi-particle BPS states in string 
theory carrying $S^3$ angular momentum. 
The agreement of the degeneracies at every level 
(up to the exception noted below) involves detailed properties of the 
dual theories and does not follow from symmetry consideration 
alone. 
Non-BPS excitations are \slt\ descendants, while 
Virasoro descendants are more general multi-particle states. 

One of the most fascinating results of the analysis is that the upper 
bound on the 
$U(1)$ charge encountered in the conformal field theory 
construction of chiral primaries translates into 
an exclusion principle limiting 
 the occupation numbers of bosonic BPS particle modes.
The maximum allowed occupation number 
grows in inverse proportion to the  
coupling constant, and is proportional to the surface area  of the 
region occupied by the particles in Planck units.
Hence the bound is nonperturbative in nature, and cannot 
be seen from perturbative string theory on \mf. 
This result is largely algebraic and 
follows in the spirit of \refs{\jbmh, \nhbh} for any sufficiently 
supersymmetric quantum theory of gravity on \ads. Very similar 
exclusion principles have been encountered in previous investigations
of nonperturbative string physics \refs{\gps,\grta, \mtrx} and 
we expect it has 
a general significance.

We also find a beautiful relationship (independent of string theory) 
between the orbifold procedure used to construct the BTZ 
black hole from \ads\ and the density matrix of the conformal field theory.
The conformal field theory lives on the cylindrical boundary of 
\ads . The boundary inherits from \ads\ a preferred set of null coordinates 
on which the \slt\ isometries have the canonical action. 
The conformal field theory 
is in the vacuum with respect to these coordinates. In general they 
differ by an exponential transformation from the coordinates 
used to define energy and momentum, and in which the 
discrete identifications act simply. This is exactly 
the transformation from two-dimensional Minkowski to Rindler 
coordinates, and the subset of \ads\ entering into the 
construction corresponds to the Rindler wedge. 
This accounts for the thermal nature of the 
mixed quantum state of the black hole. Similar relationships 
are also found between the dual  euclidean representations of the 
finite temperature partition function. 

This paper is organized as follows. In section 2 we relate the 
description of the 
lorentzian three-dimensional black hole as a quotient of 
$AdS_3$ to the conformal field theory description.
In section 3 we consider the euclidean black holes as quotients
on euclidean $AdS_3$ and relate them to similar quotients of 
conformal field theories. 
In section 3 we consider in detail the map between states
in the conformal field theory on the circle with NS boundary conditions 
and states in supergravity on $AdS_3$. 
We find the states corresponding to chiral primary fields in 
conformal field theory and discuss the bound on the particle number.
In section 4 we make some remarks about the case with Ramond boundary
conditions. 

As this work was in progress we received references 
\refs{\gkp,\ho,\withol}
which have some overlap with aspects of this paper.

\newsec{Lorentzian Black Holes}
\subsec{The Classical D1-D5 Black Hole}

In this subsection we describe the near-horizon geometry and fix our
notation. 
The low-energy action for ten-dimensional type IIB string theory
contains the terms, 
\eqn\fds{{1\over 128 \pi^7 \apm^4 }
\int d^{10}x \sqrt{- g} \[ e^{-2\phi}\(R+4(\nabla \phi )^2\)
-{1 \over 12} H^2 \]}
in the ten-dimensional string frame.
$H$ denotes the RR three form field
strength, and $\phi$ is the dilaton. The NS three form, self-dual five
form, and second scalar are set to zero.
We wish to consider a toroidal compactification to
five dimensions with an $S^1$ of length $2\pi R$ and
a $T^4$ of four-volume $(2\pi)^4 v \apm^2$.\foot{With
these conventions, T-duality sends $R$ to $\apm /R$ or $v$ to $1/v$, and S-duality
sends $g$ to $1/g$.}
In notation similar to that of \hms , the black hole solution 
associated to $Q_1$ D1-branes and $Q_5$ D5-branes is 
given in terms of the ten-dimensional
variables by
\eqn\metric{\eqalign{ e^{-2\phi }& ={1 \over g^2} \(1+   { r_5^2\over r^2 }\)
\(1 + {r_1^2 \over  r^2 }\)^{-1},\cr
H&= {2 r_5^2 \over g}  \e_3 + 2 r_1^2 g  e^{-2 \phi}  *_6 \e_3,\cr
ds^2 = &
 \( 1 + { r_1^2 \over r^2}\)^{-1/2} \( 1 + { r_5^2 \over r^2}\)^{-1/2}
 \[ - dt^2 +(dx^5)^2
\right.
\cr
+&
\left.
{r_0^2  \over r^2} (\cosh\sigma dt + \sinh \sigma  dx^5)^2
 +\( 1 + { r_1^2  \over  r^2}\) dx_i dx^i \] \cr
 +& \( 1 + { r_1^2  \over r^2}\)^{1/2}\( 
1 + { r_5^2 \over r^2}\)^{1/2} \left[
\left( 1- {r_0^2 \over r^2 } \right)^{-1} dr^2 + r^2 d \Omega_3^2 \right]
,}}
where $g$ is the
ten-dimensional string coupling, $*_6$ is the Hodge dual 
in the six dimensions $x^0,..,x^5$, the $r_i$ are functions of the mass and charges,
and $\e_3$ is the volume form on the unit three-sphere.
$x^5$ is periodically identified with period $2\pi R$,
$x^i$,  $i = 6,...,9$, are each identified with period 
$2\pi v^{1/4}\apm^{1/2}$. We are interested in the near-horizon scaling 
limit of \jm , defined by taking $\apm \to 0$ with 
\eqn\ur{U= {r \over \apm },~~~~~~~~~U_0= {r_0 \over \apm },}
as well as $v$ and $R$ fixed. In this limit \metric\ reduces to 
\eqn\tric{\eqalign{e^{-2\phi }& = { Q_5 \over g_6^2 Q_1 },\cr
{ds^2 \over \apm} =& {U^2 \over  \ell^2}(- dt^2 +(dx^5)^2)
+{U_0^2  \over \ell^2} (\cosh\sigma dt + \sinh \sigma  dx^5)^2 
 +   
{\ell^2 \over U^2-U_0^2}dU^2 \cr &+\sqrt{Q_1 \over vQ_5}dx_i dx^i +
 \ell^2 d \Omega_3^2 
,  
}}
with
\eqn\charges{
\eqalign{
   Q_1 &= {v\over 4\pi^2 g^2\apm}\int e^{2 \phi} *_6H , \cr
   Q_5 &= {1\over 4\pi^2 \apm} \int H ,
\cr
  g_6^2&={g^2 \over v},\cr
  \co &= Q_1Q_5 , \cr
 \ell^2 & = g_6 \sqrt{\co}.
}}
The shorthand $\co$ is adopted because $Q_1Q_5$ equals the 
level of the Kac-Moody superconformal algebra of the 
dual conformal field theory. 
The total momentum around the $S^1$ is given by
\eqn\mnh{{n \over R}={RU_0^2 \sinh 2\s \over 2 g_6^2}.}
All charges are normalized to be integers and taken to be positive. 
The extremal limit with non-zero momentum charge
 is $U_0 \to 0, ~\sigma \to \infty$ with 
$U_0e^\sigma$ held fixed. 

The energy above the $U_0=n=0$ black hole ground state is
\eqn\lkj{M={RU_0^2\cosh{2\s}\over 2 g_6^2}.} 
We note that this and other relevant quantities are independent 
of $\apm$, and so remain finite in the scaling limit. 
Note that the  volume of the four-torus in \tric\ is 
 $v_f = {Q_1/Q_5}$. In fact some of the scalar fields 
have fixed values depending only on the charges (the fixed scalars) \adf . 
The value of the six dimensional string coupling $g_6$, on the other
hand is not fixed and has the same value that it had asymptotically in
\metric . Similar observations apply to 
the $K3$ case.

\subsec{Black Holes as \sl\ Quotients  }
The three-dimensional part of the near-horizon geometry 
\tric\ is the BTZ black hole \sh .
In this section we review its description  as a 
discrete identification of a portion of 
$SL(2,R)$  = \ads\   \refs{\bhtz,\steif,\cang} .

The $x^5, r ,t$ part of the metric is locally \ads. To make 
this manifest define
new coordinates
\eqn\ncd{\eqalign{w^\pm&={\sqrt{U^2-U_0^2} \over U} e^{2\pi T_\pm (x^5\pm t)},\cr
y&={U_0 \over U} e^{\pi T_+(x^5+t)+\pi T_- (x^5-t)},}}
where we have introduced the left and right temperatures
\eqn\tlr{T_\pm = {1\over 2 \pi } {U_0 e^{\pm\sigma}  \over \ell^2 }. }
In terms of these coordinates the three-dimensional metric is 
locally\foot{In this and all subsequent line elements we omit for brevity an overall 
factor of $\apm$ which ultimately cancels in the string worldsheet action and drops 
out.}
\eqn\dstr{ds_3^2={ \ell^2  \over y^2}\(d y^2
+dw^+dw^-\).}
This expression is independent of $U_0$ and $\s$. 
Note that the original $x^5\pm t$ coordinates, prior to the 
periodic identification,  cover only the 
Rindler wedge of the $w^\pm$ Minkowski 
space. In the asymptotic region, $y \to 0$.
The \ads\ metric \dstr\ has an \slt\ isometry group. The \sll\ action is 
\eqn\esl{\eqalign{w^+ & \to {aw^++b \over cw^++d},\cr
                  w^- & \to w^- +{c y^2 \over cw^++d},\cr
                  y &\to {y \over c w^++d},}}
where $ad-bc=1$. The \slr\ action is similarly 
\eqn\eslr{\eqalign{w^+ & \to w^+ +{c y^2 \over cw^-+d},\cr
                  w^- & \to {aw^-+b \over cw^-+d},\cr
                  y &\to {y \over c w^-+d}.}}
Note that \esl ,\eslr\ map the boundary ($y =0$) to itself and 
act on the boundary as the usual conformal transformations of
1+1 dimensional Minkowski space.
The periodic identification of \tric
\eqn\fgh{x^5 \sim x^5+2\pi R,}
is generated by an element $P=P_L \otimes P_R$ of \slt. The action of $P$ 
on $y, w^\pm$ follows from the coordinate relation \ncd\ as 
\eqn\wrr{\eqalign{w^\pm & \sim e^{4\pi^2RT_\pm}w^\pm, \cr y&\sim 
e^{2\pi^2R(T_++T_-)}y
.}}
We can then identify $P$ as the \slt\ matrix given by
\eqn\lcm{ P_R=\pmatrix{ a & b \cr c & d }=
 \pmatrix{ e^{2\pi^2RT_+}
  &0\cr
             0& e^{-2\pi^2RT_+}
 \cr} ,}
\eqn\lycm{ P_L=\pmatrix{ 
e^{2\pi^2RT_-}  &0\cr
             0&e^{-2\pi^2RT_-} \cr }.}
$P$ is uniquely defined only up to conjugation.

The extremal limit with nonzero momentum charge is 
\eqn\xlt{U_0 \to 0,~~~~\s \to \infty , ~~~~~~{ U_0e^{\s} }
={2 g_6 \sqrt{n} \over R} ~~~{\rm fixed}.}
In this limit the coordinate transformation \ncd\ to \ads\ coordinates degenerates.
Instead the transformations 
\eqn\ncdr{\eqalign{w^+&={1 \over 2\pi T_+} e^{2\pi T_+(x^5+ t)}
,\cr
w^-&={ (x^5-t) }-{\ell^4 \pi T_+\over U^2}, \cr
y&={\ell^2  \over U} e^{\pi T_+(x^5+ t )}  }}
take the extremal metric to the \ads\ form \dstr. At $r=\infty$, we see that 
$w^-$ is the same as $x^-$, but $w^+$ is exponentially related to 
$x^+$. Hence the $x^\pm$ coordinates cover half of the $w^\pm$ Minkowski space,
and only the left movers are in a thermal state. 
The identification $x^5 \sim x^5+2\pi R$ corresponds to 
\eqn\wrrz{\eqalign{w^+ & \sim e^{4\pi^2RT_+} w^+, \cr
w^- & \sim w^-+{2 \pi R }
\cr y&\sim e^{2\pi^2RT_+} y
.}}
The corresponding \slt\ identifications for the extremal case are given by 
\eqn\zcm{ P_L=\pmatrix{e^{ 2 \pi \sqrt{n/\co}}
   &0\cr
             0& e^{ -2 \pi \sqrt{n/\co }} \cr} ,}
\eqn\zycm{ P_R=\pmatrix{ 
1  & {2 \pi R } \cr
             0 &1 .}}
We note that for small $R$ - or very low energies - the 
identification effectively acts only on the left.

The $M=0$ black hole has in addition to \xlt\ $n=0$. 
The three-dimensional part of the $M=0$ 
metric ( $U_0=0$ in \tric\ ) is transformed to the \ads\ form 
\dstr\ by the redefinitions
\eqn\rrcz{\eqalign{w^\pm &=  {x^5\pm t }, \cr
y & ={ \ell^2  \over U }.}}
The identifications are generated by 
\eqn\lfcm{ P_L=\pmatrix{1 &{ 2\pi R}\cr
             0 &1 \cr} ,}
\eqn\lfycm{ P_R=\pmatrix{1 &{ 2\pi R} \cr
             0 &1 \cr} .}
\subsec{The NS1-NS5 Black Hole}

In this subsection we consider the S-dual case of black holes 
constructed from $Q_1$ wrapped fundamental strings, $Q_5$ 
NS fivebranes and momentum. This case is of special interest 
because it is described by a known conformal field theory 
and hence much can be said about the string spectrum and dynamics. 
The S-dual of the near-horizon metric \tric ,
obtained by multiplying by $e^{-\phi}$, is  
\eqn\trc{\eqalign{d\hat s^2 
=& {\hat v U^2 \over Q_1}(- dt^2 +(dx^5)^2)
+{\hat v U_0^2  \over Q_1} (\cosh\sigma dt + \sinh \sigma  dx^5)^2 
 +   
{Q_5 \over U^2-U_0^2}dU^2 \cr &+
 d\hat x_i d \hat x^i +Q_5d \Omega_3^2 
.\cr 
}}
where $\hat v = 1/g_6^2$  is the volume of the four-torus in the dual
metric and $\hat x_i \sim \hat x_i + 2 \pi \hat v^{1/4} $.
The six dimensional dilaton is now a fixed scalar 
\eqn\dfg{{1 \over \hat v}e^{2\phi }={Q_5 \over Q_1}.} 
The coordinate transformation \ncd\ then reduces the three-metric to 
\eqn\dstl{ds_3^2={Q_5\over y^2}\(d y^2 +dw^+dw^-\),}
which corresponds to a level $Q_5$ WZW model on the 
fundamental string worldsheet and does not involve either $Q_1$ or $n$. 
The generic black hole is then constructed with the \slt 
identification \lcm\ \lycm.


\subsec{Conformal Field Theory}

In this subsection we show in the conformal field theory 
picture that the discrete identifications lead precisely to the 
density matrix expected to correspond to the black hole. 

The six-dimensional black string or five-dimensional black hole 
is described by a 
\eqn\cgh{c=6\co=6Q_1Q_5} 
conformal field theory \ascv. 
This conformal field theory can be thought of as living in 
the asymptotic boundary $r = \infty$ of the solution \tric, 
or equivalently $y = 0$ in the coordinates \dstr\ \refs{\jbmh}. 
Let us first consider 
the black string case with no $x^5$ identifications. The 
coordinates used to define energy and momentum in the CFT
are the flat coordinates in the asymptotic region of the 
full spacetime solution \metric, namely $x^5, t$.  In the generic case 
of  nonzero 
$U_0$ the conformal field theory is excited and is not in the 
$x^5,t$ vacuum. Rather it is in the \slt\ vacuum 
with respect to the $w^\pm$ coordinates defined in \ncd. 
At  $r=\infty$ these coordinates  
are related to $x^5, t$ coordinates via 
\eqn\ncdd{w^\pm= e^{2\pi T_\pm(x^5\pm t)}.}
This is precisely the relation between Rindler and Minkowski coordinates 
in two dimensions. The quantum state is the $w^\pm$ Minkowski vacuum, but 
the $x^5,t$ coordinates cover only the Rindler wedge. This is a density matrix 
rather than a pure state because of correlations with 
the quantum degrees of freedom 
behind the Rindler horizon.  
An inertial observer (moving in a straight line in $x^5, t$ coordinates) will 
detect a thermal bath of particles, with left and right temperatures given by 
$T_+$ and $T_-$. This bath leads to energy densities $T_{\pm\pm}$, 
given by the usual schwarzian transformation law
\eqn\tpmpm{T_{++}= {k  \over  2 \pi }
\sqrt{\p w^+ \over\partial x^+}
{\p^2 \over \p x^{+ 2}}
                        \sqrt{\p x^+ \over \p w^+} 
= { \pi \over 2 } \co T_+^2 .}
with $x^\pm=x^5\pm t$.
 A similar expression exists for $T_{--}$ in terms of $T_-$.
This agrees (after integrating around the circle) 
with the semiclassical expression \lkj , \tlr\
for the energy above 
extremality. 

In the extremal case, we see from \ncdr\ that $w^-$ agrees asymptotically with 
$x^5-t$ (up to a constant), but $w^+$ is exponentially related to $x^5+t$. 
Hence only the left movers are in a thermal state. For the $T_+ = T_- =0$
 black string  there are no 
exponential transformations and the conformal field theory is in its 
ground state.

\newsec{Euclidean Black Holes}

In the previous section the precise relation between lorentzian
black holes and conformal field theory density matrices, 
both expressed as \slt\ quotients, was presented. In this section 
a precise relation between euclidean partition functions will 
be found.     

\subsec{Euclidean Solutions}
In this subsection we review the euclidean BTZ black hole \scct 
 \foot{We are rescaling $t$ by a factor
of $\ell$ as compared to \scct .}.
For this purpose it is convenient to write the lorentzian near-horizon 
three-metric in \tric\ as 
\eqn\btzm{ds_L^2= -{(r^2-r_+^2)(r^2-r_-^2) \over r^2 }d t^2+{
r^2 \ell^2 \over(r^2-r_+^2)(r^2-r_-^2)}dr^2 +
r^2(d\phi +{ r_+ r_-\over  r^2}dt)^2,}
with
\eqn\rpmd{ r_\pm = {\pi   \ell}\( \tilde T_+\pm\tilde  T_-\) ~,
}
where the we have defined the R-independent temperatures 
$\tilde T_\pm=RT_\pm$.
In terms of these quantities the momentum $n$ (which 
is angular momentum from the 2+1 perspective) is 
\eqn\rfg{n={kr_-r_+ \over \ell^2} .} 
The euclidean black hole is obtained by analytic continuation of 
\btzm\ to imaginary $t$ and $J$. In terms of 
\eqn\ert{\eqalign{t&=i\tau,\cr n&= -in_E,\cr}}
the euclidean metric is 
\eqn\btze{ds_E^2= {(r^2-r_+^2)(r^2-r_-^2) \over r^2 \ell^2 }d\tau^2+{
r^2 \ell^2 \over (r^2-r_+^2)(r^2-r_-^2)}dr^2 +
r^2(d\phi + {r_+ (ir_-) \over r^2}d\tau)^2.}
For real $\tau$ and $n_E$, $ds_E^2$, $r_\pm^2$ and $r_+$ are all real,
$T_+$ and $T_-$ are complex conjugates of one another and $r_-$ is pure 
imaginary. 
The entropy is given by 
\eqn\sdt{S ={2\pi r_+ \over 4 G_3 }=2\pi^2 k  (\tilde T_+ + \tilde T_-),}
where the three-dimensional Newton's constant is $G_3=\ell /4k$.

The euclidean quantum theory expanded about \btze\ gives a contribution to 
the 
partition function for the theory at inverse temperature 
\eqn\tbh{{1 \over \tilde T }={\p S \over \p \tilde M}={1 \over 2}\bigl(
{1 \over \tilde T_+}+{1 \over \tilde T_-}\bigr).    }
Since the metric is real for imaginary $n$ there is an imaginary angular 
potential 
\eqn\ap{{ \Omega \over \tilde T} =
{\p S \over \p n}={1\over 2  }\bigl(
{1 \over \tilde T_+}-{1 \over \tilde T_-}\bigr).}
It is convenient to introduce a complex inverse temperature
\eqn\bdf{\beta={1 \over \tilde T_+}.}
\subsec{\slc }
Next we wish to represent \btze\ as  a quotient of the three dimensional
hyperbolic plane $H^3$. This 
was worked out in 
\scct. Defining new coordinates
                \eqn\eft{\eqalign{
w &= \left({r^2-r_+^2\over r^2-r_-^2}\right)^{1/2}
      \exp\left\{ {r_+ + r_-\over\ell}(\phi +{i}\tau) \right\}
      \cr
y &= \left({r_+^2-r_-^2\over r^2-r_-^2}\right)^{1/2}
      \exp\left\{ {r_+\over\ell}\phi +  {ir_-\over\ell }\tau \right\}\cr}}
the metric \btze\ takes the form
\eqn\dfr{ds^2={\ell^2 \over y^2}(dwd\bar w+dy^2),~~~~~~~ y>0.} 
The identifications implied by the thermal ensemble $ \phi + i\tau
 \sim 
 \phi + i \tau  + i \beta $ are automatically 
taken into account by the exponential map.
The periodic identification of $\phi$ implies  identifications of \dfr:
\eqn\erfg{\eqalign{w & \sim e^{4\pi^2/\beta}w ,\cr
                    y &\sim e^{2\pi^2/\beta 
+2\pi^2/\beta^*}y. } }
These can be represented as an \slc\ matrix 
\eqn\slcm{ \pmatrix{a    &b\cr
             c& d \cr} ,}
for complex $a,b,c,d$ obeying $ad-bc=1$.
The identifications \erfg\ of \dfr\ are generated by 
\eqn\drfg{H = \pmatrix{e^{2\pi^2/\beta}   &0\cr
             0& e^{-2\pi^2/\beta  }\cr} .}
\subsec{The Black Hole Partition Function}

Euclidean  partition functions are defined as 
functional integrals with fixed boundary conditions 
at infinity. In the case at hand we wish to fix the inverse
temperature and angular potential, which are the real and 
imaginary parts of $\beta$. This is implemented by the boundary 
condition that the asymptotic geometry is 
a torus with modular parameter 
\eqn\trg{\tau={ i \beta   \over  2 \pi } = {  i \over 2 \pi  \tilde T_+ } .} 
The semiclassical approximation 
to minus the logarithm of 
the partition function is then the action of the least 
action instanton obeying these boundary conditions. 

The euclidean black hole with $\beta$ given by \bdf\ provides such 
an instanton. The action as a function of $\beta$ is 
\eqn\bha{S_{instanton}=-{\pi^2 k \over  \beta }-{\pi^2 k \over  \bar \beta}
=- {i\pi k \over 2}({1 \over \tau} -{1 \over \bar \tau}) .}
A second instanton corresponds to a thermal gas in \ads . It is 
constructed as the periodic identification in euclidean time 
(denoted $\tau_E$ in this subsection 
to avoid confusion with the modular parameter \trg )
\eqn\ident{\tau_E + i \phi  \sim \tau_E+i\phi +\beta } 
of the metric \btze\ with $r_+^2=-1$ and $r_-=0$ 
( namely euclidean \ads ):
\eqn\dfrh{ds^2={( r^2+\ell^2 )}d\tau_E^2+ \ell^2  
{dr^2 \over r^2+ \ell^2}+r^2d\phi^2. }
  From 
the action of the isometries in this coordinate system 
it can be seen that \ident\ is generated by 
the \slc\ element
\eqn\drg{H = \pmatrix{e^{i\beta/2}   &0\cr
             0& e^{-i\beta/2 }\cr} .}
The action of this instanton is the (negative) mass of \ads\ times the 
euclidean time:
\eqn\dcv{S_{instanton}=-{k (\beta +\bar \beta)\over 4}
={i\pi k \over 2}({ \tau}-{ \bar \tau}).}

Notice that \bha\ and \dcv\ are related by the $SL(2,Z)$ transformation
$\tau \to -{1 \over \tau}$. One might accordingly anticipate an $SL(2,Z)$ 
family of 
instantons with actions
\eqn\sgr{S_{instanton}
={i\pi k \over 2}({a\tau +b\over c\tau +d}-
{a\bar \tau +b\over c\bar \tau +d}).}
These can be constructed beginning with the \slc\ identification
\eqn\drcg{H = \pmatrix{e^{i\pi{a\tau +b\over c\tau +d}}   &0\cr
             0& e^{ -i\pi{a\tau +b\over c\tau +d} }\cr} .}
Comparing with \drfg\ we see that this yields an instanton whose 
asymptotic torus has modular parameter
${a\tau +b\over c\tau +d}$ in the coordinates \dfr. 
However this modular parameter can be transformed to 
$\tau$ by an $SL(2,Z)$ transformation of the coordinate $\tau_E+i\phi$.
Hence for every identification of the form \drcg\ 
there is a way of identifying 
the asymptotic torus so that the geometry satisfies the boundary 
conditions imposed in the functional representation of the partition 
function at complex inverse temperature $\beta$.  

In each modular region of the $\tau$ plane there is a unique 
lowest (negative) action instanton. Defining
\eqn\ytg{S_{min} (\tau) \equiv  {\rm min} 
\[{i\pi k \over 2}({a\tau +b\over c\tau +d}-
{a\bar \tau +b\over c\bar \tau +d}) \] ,}
the leading semiclassical approximation to the partition function is 
\eqn\jil{Z(\tau)=e^{-S_{min}(\tau)}.} 
At low temperatures ( large $\beta$ ) the partition function is dominated 
by \dcv\ corresponding to a thermal gas in \ads . At higher temperatures 
there is a transition to the black hole 
phase in which \bha\ dominates.  This is a sharp first order phase transition 
in the limit $ k \to \infty$ \withol . 

\subsec{Conformal Field Theory Partition Function}
The euclidean conformal field theory lives on the 
complex plane with coordinates $(z, \bar z)$. 
The action of the global \slc\ on $(z, \bar z)$ is 
\eqn\dddg{z \to {az+b \over cz+d},}
for complex $a,b,c,d$ obeying $ad-bc=1$. 
$H^3$ corresponds to the \slc\ invariant vacuum on the plane. 
Hence the 
black hole partition function should be equivalent to 
a conformal field theory partition function on the plane 
identified as 
\eqn\zidf{z \sim e^{4\pi^2/\beta}z.}
This is a toroidal partition function, with modular parameter
$\tau=i2\pi  / \beta  $. It is represented in radial 
quantization as an annulus 
with inner and outer edges glued together. 
Note that the fermions are periodic around the radial cycle and 
antiperiodic around the angular cycle.

To interpret this as a finite-temperature 
conformal field theory partition function, 
consider a modular transformation  $\tau \to -1/\tau$.
One then has a torus with modular parameter 
$\tau= i \beta/2\pi $ . The antiperiodic direction has length
$ 1/\tilde T$, as in the standard representation of 
a thermal partition function at temperature $\tilde T$. The imaginary 
part of $\tau$ leads to the anticipated phase of $-\Omega n /\tilde  T$ for a 
state with $n=n_L-n_R$.
We see that there is a precise correspondence between  
the element of the \slc\ isometry group of $H^3$ used to construct the 
euclidean black hole and the element of the \slc\ conformal group of the 
plane used to construct the conformal field theory partition function. 
Notice that the supergravity result \ytg\ implies that for large $k$ 
only the 
vacuum contributes once we look at the system in the appropriate 
channel. For example consider the case with $\Omega =0$ then we have
the CFT on the rectangle. The supergravity result \ytg\
means  that once we take as euclidean time the longest
side of rectangle, only the vacuum propagates. 


\newsec{The \ads\ $\leftrightarrow$ Conformal Field Theory Map}

In this section we study the correspondence between states in $AdS$ and
in the conformal field theory. We first start with some general properties
which follow from conformal invariance and then we study properties
which are more specific to the (4,4) conformal field theories corresponding
to type IIB string theory on $AdS_3\times S^3\times M^4$,
where $M^4$ is $K3$ or $T^4$.

\subsec{The NS-NS Sector}

In this section we consider the NS-NS sector with antiperiodic fermions, 
which turns out to be the simplest case, i.e. we consider the
conformal field theory on a circle of radius one with NS boundary
conditions
for the fermions. This is the state that we get if we map the plane to
cylinder
without inserting any operator at the origin. 

In the spacetime picture the NS-NS vacuum corresponds simply to the 
\ads\ vaccum defined with respect to the coordinates 
\refs{\ch, \chtwo }
\eqn\mtric{
{ds^2 \over \ell^2} = - \cosh^2 \rho d\tau^2 + \sinh^2 \rho d\phi^2 + d \rho^2
}
The fermions on the cylinder at infinity 
are antiperiodic because $\phi \to \phi+2\pi$ is a $2\pi$ rotation 
of the constant-time disc.
In contrast the RR vacuum, where the fermions are periodic
under $\phi \to \phi+2\pi$, 
 corresponds to  the $M=0$ black hole where
the surfaces of constant time are not a disc.
The three-dimensional mass of \ads, defined with respect to 
the RR vacuum, is  $M_3=-{c \over 12}$.
This is the value expected from the usual conformal field 
theoretic mass shift formula 
for the NS-NS vacuum. 
In the conformal field theory, NS-NS states are in one-to-one correspondence 
with local operators on the plane. 
The coordinate transformation mapping \mtric\ to \dstr, given in section 
5 below,  reduces 
to the map between the plane and (a diamond in) the cylinder on the boundary. 
Hence we expect a map between 
the string Hilbert space on \mf\ and operators of the conformal field theory 
on the plane.

The first step is to identify the \slt\ representations of the 
states/operators. 
In the coordinates \mtric , with $u = \tau + \phi$, $v = \tau-\phi$, 
the $SL(2,R)_L$ generators are described by the vector fields
\eqn\genleft{
\eqalign{
L_0 = &  i \partial_u ~, \cr
L_{-1} = &  i e^{-iu} \left[ { \cosh 2 \rho \over \sinh 2 \rho } \partial_u
- { 1 \over \sinh 2 \rho} \partial_v +  { i \over 2} \partial_\rho 
\right] ~, \cr
L_{1} = &  i e^{iu} \left[ { \cosh 2 \rho \over \sinh 2 \rho } \partial_u
- { 1 \over \sinh 2 \rho} \partial_v -  { i \over 2} \partial_\rho  
\right]~, }}
normalized so that
\eqn\commut{
[L_0,L_{\pm1} ] = \mp  L_{\pm 1}~,~~~~~~~~~~~~[L_1,L_{-1} ] = 2 L_0~. 
}
The $SL(2,R)_R$ generators $\bar L_0, \bar L_{\pm1}$ 
are given by a similar expression with $u \leftrightarrow v$.
The quadratic Casimir of $SL(2,R)_L$ is  
\eqn\casim{
L^2 = { 1\over 2} ( L_{1} L_{-1} + L_{-1} L_1 ) - L_0^2 ~.
}
The sum of the two Casimirs is 
\eqn\slum{
- 2 ( L^2 + \bar L^2 ) = \partial_\rho^2 + 2 { \cosh 2\rho \over
\sinh 2 \rho } \partial_\rho  + {1 \over \sinh^2 \rho} \partial^2_\phi
- { 1 \over \cosh^2 \rho } \partial_\tau^2 ~,
}
which is the laplacian on scalar fields times $\ell^2$. Therefore a scalar field of mass 
$m$ has 
\eqn\mlk{L^2 + \bar L^2=-m^2 \ell^2/2 ~,}
and the conformal algebra can be used to classify the solutions of
the wave equation. 
In the usual manner consider states with weights $(h,\bar h)$ under
$L_0,\bar L_0$ so that
\eqn\eigenv{
L_0 | \psi \rangle = h  | \psi \rangle ~,~~~~~~~~~~
\bar L_0 | \psi \rangle = \bar h  | \psi \rangle ~.
}
It follows that 
\eqn\wave{
| \psi \rangle = e^{- i h u  -i \bar h v } F(\rho) ~.
}
Now suppose that $| \psi \rangle$ 
is a primary state in the
sense that $L_1 | \psi \rangle = \bar L_1 | \psi \rangle = 0$.
These conditions imply that
that $ h = \bar h$ and that $F$ satisfies
\eqn\eqnprim{
2 h { \sinh \rho \over \cosh \rho }F +  \partial_\rho F  =0 ~,
} 
which is solved by 
\eqn\soln{
F = const { 1 \over ( \cosh \rho )^{2 h} } ~.
}
Demanding that $ | \psi \rangle$ represents a scalar of mass $m$ 
imposes the additional constraint \mlk .
Since $L^2 = L_{-1} L_1 - L_0(L_0 -1)$ we find that this condition implies
that the normalizable solutions have 
\eqn\frg{h = \bar h = {1 \over 2}\left(1  +
 \sqrt{{m^2 \ell^2 }+ 1} \right) ~.}
Note that massive string states have $m^2$ of order one and 
hence correspond to very large $L_0$ eigenvalues.  
For a massless scalar particle in the $l$th partial wave on 
 $S^3$ one finds
that $h =\bar h= 1 + l/2 $. The $l$th partial wave
transforms in the  ($l/2$,$l/2$) representation of  $SU(2)_L \times
SU(2)_R$. 
As explained in \refs{\stabil} stability analysis on $AdS_3$ 
requires that that $m^2 \ell^2 \geq -1 $ so that the dimension
of the corresponding primary operator is $h \geq 1/2$.

Starting from the primary state \soln\ we can generate all other
normalizable solutions with Dirichlet boundary conditions at
infinity by acting with $L_{-1}, ~ \bar L_{-1}$, 
since starting from any state we can act with $L_1,~ \bar L_1$ and
lower its energy. Since the energy should be positive for the
state to have positive norm we conclude that at some point we
will get a primary field, and the primary field \soln\ is unique. 
%
These \slt\ descendants all have the same quadratic Casimirs,
but higher integral eigenvalues of $L_0,\bar L_0$.
These states correspond in an obvious way to \slt\ descendents 
of the primary operators.
In particular for a massless scalar $h=\bar h =1$,
and the primary field corresponds to a $(1,1)$ operator. Each modulus
of the conformal field theory should therefore correspond to a massless
scalar in $AdS_3$. We shall see below that this is indeed the case. 
Conformal field theory operators may be further classified by their 
$R$-charges which identifies the $S^3$ angular momentum of the corresponding 
state. 

\subsec{Chiral Primaries}

The conformal field theory that we are considering has (4,4) supersymmetry.
 A particularly interesting set of operators are the 
chiral primaries, for which the preceding considerations lead to a unique
identification with a state in \mf . 
The conformal field theory that we are dealing with is some deformation
of $Sym_{\co} M^4$. 
To be specific let us consider 
in the conformal 
field theory the (a,c) ring with 
respect to some 
particular N=2 subalgebra for the $K3$ case.
In order to compute the chiral ring we can  take the theory to
be simply the orbifold $Sym_{\co} M^4$ since it does not depend on
the moduli of the conformal field theory.  
 For $\qof =1$, the conformal 
field theory target space is a single 
copy of $K3$, and the chiral primary fields are 
constructed from the 24 harmonic $(p,q)$ forms $\omega^A$, 
for $A=1,...24$   as 
\eqn\sddfg{ \Phi^A=\omega^A_{ab..\bar a \bar b ..}\psi^a (z) 
\psi^b(z)..\psi^{\bar a} (\bar z) \psi^{\bar b}(\bar z)...}
This set of fields includes the identity. The $R$ charges 
are just the rank $(p,q)$ which corresponds to twice the 
spacetime angular momentum.  The cohomology of 
$Sym_\co (K3)$ is the $\qof$-fold product of the $K3$ 
cohomology, plus twisted sector contributions from the 
fixed points of the permutation group \vw . Let us first consider 
the subset invariant under 
the permutation group $S_\qof$. This can be described by 
introducing one species of bosonic creation operator
$\alpha^{A}_{-1}$ for each $K3$ cohomology class\foot{
For the $T^4$ case we have to include also fermionic oscillators
corresponding to odd cohomology classes.}.  The space of symmetric 
cohomology elements of $(K3)^\qof$ are  then isomorphic to 
the $\qof$ particle Hilbert space. When twisted sectors are included,
one has additional creation operators $\alpha^{A}_{-n},~~1 \le n \le \qof$
and considers the Hilbert space at level $\qof$. Hence a general 
operator involves $M \le \co $ oscillators, and 
can be written
\eqn\gst{\prod_{i=1}^M \alpha^{A_i}_{-n_i}| 0 \rangle ,}
where $\sum n_i=\co$. 
With each oscillator one can associate the rank $(p_i, q_i)$ of its corresponding 
cohomology element. 
The $R$ charge of the operator \gst\ is then 
\eqn\rfg{ (P,Q)= (\co -M + \sum p_i, \co -M +\sum q_i).}
Since these are chiral primaries one also has 
$(h, \bar h)=({P \over 2}, {Q \over 2})$.

We now identify the corresponding states of IIB string theory, 
beginning with the states with small charges. 
The vacuum corresponds to the unique state with  $R$-charge $(0,0)$,
 constructed from $\co$ level one identity oscillators. 

At charge $(1,1)$ there are 20 operators constructed 
from $\co -1$ level one identity oscillators and one level one 
rank $(1,1)$ oscillator. There is an additional operator from 
$\co -2$ level one and one  level two identity oscillators. 
This gives a total of 21 charge $(1,1)$ oscillators. 
Since $h=\bar h = \half$ we see from \frg\ that these 
operators should correspond to states with negative mass 
$m^2=-{1 \over \ell^2}$ on \ads. 

Now we turn to the supergravity  description of the 
corresponding states. 
Compactifying IIB supergravity on $K3$ gives the unique 
anomaly-free six-dimensional $(0,2)$ theory with 21 anti-self-dual 
and 5 self-dual
rank three antisymmetric tensor field strengths. Further compactification
on $S^3 \times$\ads\ requires an expectation value for one of the 
self-dual tensors. 
In the six dimensional theory we have 5$\times$21 scalars living in 
$SO(5,21)/SO(5)\times SO(21)$ which are in the same supersymmetry 
multiplet as the anti-self-dual field strengths. When we compactify
on $AdS_3\times S^3$ 21 of the scalars acquire fixed expectation
values and their fluctuations around this value get a mass. The 
other 4$\times$ 21 scalars are still massless fields. We define
the mass as the number appearing in \mlk . All these are scalars 
already in six dimensions. In addition, we will have scalar
fields on $AdS_3$ coming from the Kaluza-Klein reduction on 
the sphere. In particular, 
the above considerations lead us to expect  21 scalar fields
on $AdS_3$ with $m^2 \ell^2 = -1$ corresponding to the 
minimal weight chiral primaries. 
We shall see below that these fields arise from the Kaluza-Klein 
reduction of the anti-self-dual field strengths on $S^3$.
In the process we get, as a by product, the 21 fixed scalars.
Let us denote by $H^1$ the self dual field strength which is nonzero.
All scalar fields 
are constant and can be rotated by a global SO(5,21) symmetry
so that we now expand around the identity element in the coset
$SO(5,21)/SO(5)\times SO(21)$.
 For small fluctuations we can think of the
scalars as a 5$\times$21 matrix $w^{im}$, $m=1,..,21$.
The $4\times 21$ scalars $w^{im}$, $i\not = 1$ do not 
mix with anything and lead to massless fields.
 We denote the anti-dual-field strengths by  $K^m$.
For $K^m$ and $w^{1m}$ we  get the following equations \romans\
 (the hatted indices run from 0 to 5) 
\eqn\ksed{
K^m = - * K^m}
\eqn\exprk{
K^m_{\hat \mu\hat\nu\hat\rho} = w^{1m} H^1_{\hat \mu\hat\nu\hat\rho}
  +F^m_{\hat \mu\hat\nu\hat\rho}~,~~~~~~~~~F^m = d B^m~,
}
where $B^m$ is a two form.
We also get the equation
\eqn\scalar{
(\nabla_x^2  + \nabla_\theta^2) w^{1m} - { 2 \over 3} 
H^{1\hat \mu\hat\nu\hat\rho} 
K_{\hat \mu\hat\nu\hat\rho}^m =0~.
}
where $\nabla_x^2$  and $\nabla_\theta^2$ are the laplacian on \ads\ and
$S^3$ respectively.
Since the equations are diagonal in the index $m$ we drop it from now on.
We can rewrite \ksed\  with \exprk\ as
\eqn\sdcond{\eqalign{
 &F_{\mu\nu\rho}  + { 1 \over 6 }  \epsilon_{\mu\nu\rho} 
 \epsilon^{ijk} F_{ijk} 
+ 
{ {2} \over \ell } w  \epsilon_{\mu\nu\rho}  =0 ~,
\cr  
&F_{ij\mu} + { 1 \over 2 } \epsilon_\mu^{~ \nu\rho} 
\epsilon_{ij}^{~~ k}  F_{k\nu\rho}
=0~,
}}
where we 
separate the indices on $AdS_3$ (greek) and $S^3$ (latin).
We have also used  that $ \ell 
 H_{\mu\nu\rho}  =  \epsilon_{\mu\nu\rho}$, $\ell H_{ijk} =
\epsilon_{ijk}$
and all other components are zero, as   
follows from the zeroth order Einstein equations.  
Now we write the following expansions for $B$, after imposing the gauge
fixing conditions $D^i B_{i\mu} = D^iB_{ij} =0$,
\eqn\expa{\eqalign{
B_{\mu\nu} = & b_{\mu\nu}^{I_1} Y^{I_1}~, \cr
B_{\mu i} = & b_\mu^{I_3} Y_i^{I_3} ~,\cr
B_{ij} = & b^{I_1} \epsilon_{ij}^{~ \ k} D_k Y^{I_1} ~,
}}
where we have used scalar spherical harmonics and vector spherical harmonics.
Plugging these equations into the second  equation in \sdcond\ yields
an equation for $b_\mu$ which decouples from the rest of the 
equations.  From the same equation we further conclude  
\eqn\bmunu{
b_{\mu\nu}^{I_1} = \epsilon_{\mu\nu}^{~~ \ \rho} \partial_\rho b^{I_1} ~,
}
if $l>0$. 
If $l=0$ then $b_{\mu\nu} = b =0$. 
Using this equation the first  equation in \sdcond\ becomes
\eqn\bandw{
\ell^2 \nabla^2_x b^{I_1} - { C(I_1) }  b^{I_1} + 
{ 2 }  w^{I_1} =0~,
}
where $C(I_1) = l(l+2) $ is the eigenvalue of the Laplacian on the unit
three-sphere
for the correponding spherical harmonic (and $l=0,1,2\cdots $).
 and we have also expanded $w$ in 
spherical harmonics.
Using \bandw\ we see that \scalar\ becomes
\eqn\scaltwo{
\ell^2 \nabla_x^2 w^{I_1} -C(I_1) w^{I_1} - 8  
(-  C(I_1) b^{I_1} +  w^{I_1}) =0 ~.
}
The two mass eigenvalues are 
\eqn\masseig{
\ell^2 m^2_- = l(l-2)~,~~~~~~~~~ \ell^2 m^2_+ = l(l+6) + 8~.
}
Using \frg\ the left and right moving weights are 
\eqn\confw{\eqalign{
\Delta_- =& l/2~,~~~~~~~~~~l>0~, \cr
\Delta_+ = & (l+4)/2 ~,~~~~~~~l\geq 0~.
}}
We have included the result for $l=0$ which involves only  
a mode of $w$. 
The first branch in \confw\ describes  the scalar corresponding
to the chiral primary operator and all its partial waves. The
second branch corresponds to the fixed scalar.

Hence we conclude that the 21 charge $(1,1)$ chiral primaries 
correspond to a single quantum in one of the lowest-lying modes of 
the 21 $\ell^2 m^2  =-1$ scalar fields. Excited single particle 
states can be constructed
as \slt\ descendants.  

Next let us consider the charge $(n,n)$ chiral primaries. 
There are 20 such operators from $\co -n+1$ level one identity 
oscillators and one rank $(1,1)$ level $n$ oscillators,
as well as one from $\co-n$ level one identity oscillators and 
one level $n+1$ identity oscillator. These can be identified as 
single particle states arising from higher $S^3$ angular harmonics 
of the 21 self-dual tensor multiplets \confw . 

In addition for $n \ge 2$ there is a charge $(n,n)$ operator from  
$\co -n +2$ level one identity oscillators and one rank $(2,2)$ 
level $n-1$ oscillator.  The lowest state corresponds
to an operator with angular momentum $(1,1)$ under 
  $SU(2)_L\times SU(2)_R $  and therefore conformal weight (1,1). 
It could come from deformations of the sphere but
we have not checked this explicitly. 
We also expect chiral primaries coming from the $(0,2)$ and $(2,0)$ forms. 
These correspond in the conformal field theory to the 
positively-charged $SU(2)_L$ and $SU(2)_R$ currents
$J_L^+(z)$ and $J_R^+(z)$. In supergravity we have an $SU(2)_L\times
SU(2)_R $
gauge symmetry coming
from isometries of $S^3$. The positively-charged components 
of the gauge boson  correspond 
to these chiral primary states.

Finally at charge $(n,n)$ there are additional operators 
which involve multiple  oscillators that are not level 
one identity operators. These correspond in the obvious manner to 
multi-particle states, where the total particle number is the 
number of oscillators that are not level one identity oscillators.

\subsec{The Stringy Exclusion Principle}

For ordinary particle numbers, we saw in the last subsection that 
the spectrum of chiral primary operators agrees with the expected 
BPS supergravity spectrum. However it is abruptly terminated when 
the particle number reaches $2\qof$. This follows from  
fermi statistics of the $2\qof$ fermionic 
operators used in the construction of the chiral primaries. 
In  general  the 
maximum R-charge for a chiral primary is bounded by unitarity as 
$Q \leq c/3  = 2 k$. 
This effect can not be 
seen in IIB-\ads\ perturbation theory. From that point of view they 
are free bosons and there would not seem to be a limit on 
the occupation number of any particular mode. 

It is worth emphasizing that the validity of the 
bound follows from general symmetry considerations and 
does not depend on string theory. As discussed in \refs{\jbmh, \ch, \nhbh} 
any consistent quantum theory of gravity on \ads\ 
is a conformal field theory.  If it has in addition sufficiently many 
spacetime supersymmetries it will be a $(2,2)$ superconformal theory.
In general the lowest-charge (except the identity) chiral primary 
of the conformal field theory will be a single-particle state on 
\ads .  A second chiral primary can in general be constructed 
by squaring this chiral primary. This operator (when it is non-zero) 
corresponds to two quanta in the same mode on \ads. In general the 
$N$th power of the chiral primary corresponds to $N$ particles 
in the same mode. However as mentioned above it follows from the 
${\cal N}=2$ superconformal 
algebra that the $U(1)$ charge of a chiral primary is bounded  
by $\co$. Hence for $N$ of order $c$ the chiral primary 
will vanish, in accord with the exclusion principle. 

Similar phenomena have been encountered in other 
string theoretic contexts. For example  consider the 
formulation of two-dimensional $SU(N)$ gauge theory on a circle as 
a string theory \grta . This theory has states in which the string winds 
$m$ times around the circle. However one must have $m <N$ because 
$N$ powers of the gauge theory flux tube are trivial. This constraint
is invisible in the string perturbation theory. Also in the fermionic 
formulation of the $c=1$ matrix model \mtrx, excitations of the upper 
branch of the Fermi sea are abruptly cut off when the lower 
Fermi sea is encountered. This again is a bound on the number of 
quanta in a single mode.
In the case of the (0,2) six dimensional CFT 
 it was found in \ofer\ that 
the single particle spectrum of chiral primaries is bounded by the number
of branes, and a similar effects occurs for ${\cal N} =4$ $U(N)$ 
super-Yang-Mills since the trace of more than $N$ matrices can be
decomposed
as products of traces. However in those cases there does not seem to 
be a bound for multiparticle states. Closely related observations 
of a bound on $S^3$ angular momentum in the context of \ads\ 
scattering were made in \gps.

\subsec{The Virasoro Generators}

The primary operators of the conformal field theory 
have Virasoro as well as \slt\ descendants. 
These descendants in general have differing \slt\ Casimirs and 
correspond to multi-particle states on \ads. In this 
subsection we discuss the 
Virasoro action on \ads. 

In quantum gravity the canonical generators of diffeomorphisms contain
volume integrals of the constraints. These integrals must be augmented  by 
surface terms in order to have the correct algebra.  
The latter in turn can be expressed as  the volume 
integral of a total divergence. After fixing the gauge appropriately 
and imposing the 
constraints the  Virasoro generators $L_n$ on \ads\ 
can  be written as the volume integral \jbmh\ 
\eqn\vgn{L_n=\int d\Sigma^\mu \zeta^\nu_n \hat T_{\mu \nu}.}
$\hat T$ is the matter stress tensor plus a gravitational 
stress tensor given by the Einstein tensor minus the 
total divergence of the surface term\foot{Since there are no 
gravitons in three dimensions this could vanish in an appropriate gauge.}.
The vector fields $\zeta_n$  
\eqn\zta{
\eqalign{\zeta^+_n&= e^{in u^+},\cr
                 \zeta^-_n&= {n^2 \over r^2}e^{in u^+},\cr
                 \zeta^r_n&=-{ irn \over 2} e^{in u^+}.}}
The choice of subleading in ${1 \over r}$ corrections is ambiguous, 
as such terms manifestly do not contribute in 
the boundary expression for $L_n$. We have chosen these terms so that
the Lie derivative with respect to $\zeta_n$ obeys
\eqn\dfc
{\[
{\cal L}_{\zeta_m },{\cal L}_{\zeta_n }\]
=i(n-m){\cal L}_{\zeta_{m+n}},}
although a central charge of course appears in the Poisson brackets
of the generators \vgn . 
$\bar L_n$ is given by a similar expression with  
$+\leftrightarrow -$. The \slt\ subgroup 
generated by $L_0,~L_{\pm 1},~\bar L_0,~\bar L_{\pm 1}$  is the isometry
group of \ads\ ( although the extension of the vector field into the 
interior differs from \genleft\ at subleading order).

The action of the Virasoro generators on a quantum state  will mix up 
states with different particle numbers, because the vacuum itself 
cannot be annihilated by all the Virasoro generators. In general the 
vacuum state depends on the coordinate choice used to distinguish 
positive and negative frequency excitations. The Virasoro generators 
transform 
the vacuum defined with respect to a particular coordinate system 
to a new state 
which is the vacuum with respect to the new coordinates. 

\newsec{The RR Sector}

The preceding sections have concentrated on the NS-NS 
sector of the conformal field theory. 
In this section we first review the map between the cylinder and
the plane and then we make a few remarks about the RR sector. 

The map between the lorentzian cylinder \mtric\  and the plane \dstr\ 
is given by
\eqn\map{ \eqalign{
{ 1 \over y }  &= \cosh \rho \cos \tau + \sinh \rho \cos \phi 
\cr
t &=  y \cosh \rho \sin  \tau 
\cr
x &=  y \sinh \rho \sin \phi 
}}
where $ x \pm t = w^\pm $.
Remembering that $u=\tau + \phi $ and $v = \tau - \phi$ we 
see that at the boundary $y = 0^+$ this change of coordinates becomes
$w^+ = \tan {u \over 2 } $, $ w^- = - \tan { v \over 2 } $,
 which 
is the change of coordinates from the plane to (a diamond in) the cylinder.

The $SL(2,R)_L$ generators on the plane are 
\eqn\virplane{\eqalign{
 H_{-1} = &  i \partial_+ 
\cr
 H_{0}  = & i ( w^+ \partial_+ + {1\over 2} y\partial_y )
\cr
 H_1 = & i ( (w^+)^2 \partial_+  +  w^+ y \partial_y - y^2 \partial_- )
}}
And similarly for $SL(2,R)_R$ in terms of 
$  {\bar H}$ with $+ \leftrightarrow -$. 
Notice that the generators in \virplane\ are related to those in 
\genleft\ by
\eqn\relation{
H_0  = { 1 \over 2 i} ( L_1 - L_{-1} ) ~~~~~
H_{\pm 1} = L_0 \mp { 1 \over 2} ( L_1 +  L_{-1} ) 
}
and the scalar wave operator is 
\eqn\waveplane{
-2 (L^2 + \bar L^2 ) =  y^2 ( 4 \partial_+ \partial_- + \partial_y^2 -
{ 1\over y} \partial_y  )
}
Plane wave solutions  for the massless scalar field
equation which diagonalize $H_{-1}, 
\bar H_{-1}$
are Bessel functions of the form
\eqn\solbess{
\Phi = e^{ -i \omega t + i p x } y J_{1} ( \sqrt{ \omega^2 - p^2 } y)
}
where we have imposed Dirichlet boundary conditions  
at $y= 0 $ ($U = \infty$).
The RR sector is obtained by compactifying as indicated in \lfcm \lfycm .
This  is a transformation generated by $H_{-1} +  \bar H_{-1}$ which is 
the momentum along the plane. Notice that now the fermions
are naturally periodic around the circle. 
 As mentioned above this gives
the $M=0$ BTZ black hole which has a singularity at $ y = \infty$.
This is the location of the horizon in the non-compactified picture
(with infinitely long D1-D5 branes). The singularity appears because 
the circle along which we are identifying becomes null (of zero size)
when we approach the horizon ($y = \infty $). The 
boundary ($y = 0$) is a cylinder as in the NS-NS case. 

The singularity at $y=\infty$ makes the Hilbert space difficult to analyze.  
As opposed to the NS-NS case now the quantum gravity vacuum 
in this geometry is expected to be highly degenerate. In fact 
from the conformal field theory picture we learn that the 
asymptotic degeneracy should go as $ e^{ 2\pi \sqrt{ c_e k /6 }}$ where
$c_e = 24 $ for $K3$ and $c_e = 12 $ for $T^4$ \vafaud . 
(These degeneracies are predicted by U-duality from 
the degeneracy of perturbative string states in heterotic and
type II string theory respectively). Furthermore we expect to 
have an energy gap of the order $\Delta \omega \sim 1/k $ \refs{\msu}.
Note that if we were to look for states in this theory 
corresponding to scalar fields, as we did for the NS-NS sector,
we would find an infinite number of states corresponding to 
solutions of \solbess\ with $p=0$ and any energy $\omega$, obtained by 
continuous rescalings. 
This is not inconsistent since indeed the gap is zero 
to leading order in $1/k$. It would be interesting however to 
understand more precisely in the gravity picture the origin of
this gap.

It would be interesting to extend this analysis to other cases
 involving \ads\  described in \refs{\jm,\skend}.

\centerline{\bf Acknowledgments}

We would like to thank H. Ooguri, J. Schwarz and C. Vafa for discussions.
We would also like to thank David Gross and the
 Institute for Theoretical Physics
at the University of California at Santa Barbara, where part of 
this work was done, for hospitality.  
This work was supported in part by DOE grant DE-FG02-96ER40559.

\listrefs

\bye